\documentclass[english]{IEEEtran}
\usepackage[LGR,T1]{fontenc}
\usepackage[latin9]{inputenc}
\usepackage{float}
\usepackage{amsthm}
\usepackage{amsmath}
\usepackage{amssymb}
\usepackage{graphicx}

\makeatletter

\DeclareRobustCommand{\greektext}{%
  \fontencoding{LGR}\selectfont\def\encodingdefault{LGR}}
\DeclareRobustCommand{\textgreek}[1]{\leavevmode{\greektext #1}}
\DeclareFontEncoding{LGR}{}{}
\DeclareTextSymbol{\~}{LGR}{126}
\providecommand{\tabularnewline}{\\}

\theoremstyle{plain}
\newtheorem{thm}{\protect\theoremname}
\theoremstyle{plain}
\newtheorem{lem}[thm]{\protect\lemmaname}
\theoremstyle{remark}
\newtheorem{rem}[thm]{\protect\remarkname}
\theoremstyle{plain}
\newtheorem{prop}[thm]{\protect\propositionname}

\usepackage{cite}
\usepackage{url}

\@ifundefined{showcaptionsetup}{}{%
 \PassOptionsToPackage{caption=false}{subfig}}
\usepackage{subfig}
\makeatother

\usepackage{babel}
\providecommand{\lemmaname}{Lemma}
\providecommand{\propositionname}{Proposition}
\providecommand{\remarkname}{Remark}
\providecommand{\theoremname}{Theorem}

\begin{document}
\global\long\def\hdist#1{\mathsf{d}_{H}^{#1}}

\global\long\def\tdist#1{\mathsf{d}_{\circ}^{#1}}

\global\long\def\eqclass#1{\mathsf{R}_{#1}}

\global\long\def\eqv#1{\equiv_{#1}}

\global\long\def\noteqv#1{\not\eqv{#1}}

\global\long\def\mpcode#1{\mathsf{MPC}{}_{#1}}

\global\long\def\pcode#1{\mathsf{PC}{}_{#1}}

\global\long\def\ms#1#2{\mathsf{M}(#1,#2)}

\global\long\def\mlp#1#2{\mathfrak{m}_{#2}^{#1}}

\global\long\def\mp{\mathfrak{m}}

\global\long\def\Mlp#1#2{\mathfrak{M}_{#1}(#2)}

\global\long\def\Mp{\mathfrak{M}}

\global\long\def\orp#1#2{\mathfrak{o}_{#2}^{#1}}

\global\long\def\op{\mathfrak{o}}

\global\long\def\Orp#1#2{\mathfrak{O}_{#1}(#2)}

\global\long\def\Op{\mathfrak{O}}

\global\long\def\lcs#1#2{\mathsf{LCS}(#1,#2)}

\global\long\def\ball#1#2{\mathfrak{\mathsf{B}}_{#2}^{#1}}

\global\long\def\proj#1#2{#1_{#2}}

\global\long\def\closestpair#1#2#3{\mathsf{U}_{#1}^{*}(#2,#3)}

\title{Multipermutation Codes in the Ulam Metric\\
for Nonvolatile Memories}

\author{Farzad Farnoud (Hassanzadeh) and Olgica Milenkovic \\
Department of Electrical and Computer Engineering\\
 University of Illinois at Urbana-Champaign}
\maketitle
\begin{abstract}
We address the problem of multipermutation code design in the Ulam
metric for novel storage applications. Multipermutation codes are
suitable for flash memory where cell charges may share the same rank.
Changes in the charges of cells manifest themselves as errors whose
effects on the retrieved signal may be measured via the Ulam distance.
As part of our analysis, we study multipermutation codes in the Hamming
metric, known as constant composition codes. We then present bounds
on the size of multipermutation codes and their capacity, for both
the Ulam and the Hamming metrics. Finally, we present constructions
and accompanying decoders for multipermutation codes in the Ulam metric.
\end{abstract}

\section{Introduction}

Permutations and multipermutations as information representation formats
have a long history, with early applications in communication theory
dating back to the work of Slepian~\cite{slepian1965permutationModulation},
who proposed using multipermutation codes for transmission in the
presence of additive white Gaussian noise. More recently, Vinck proposed
using permutation codes in the Hamming metric for combatting impulse
noise and permanent frequency noise in power grids~\cite{vinckCodedModulation}.
Permutation codes have received renewed interest in the past few years
due to their promising application in storage systems, such as flash
memories~\cite{bruck2009rank-modulation,barg2012codes,engad2011constantWeightGrayCodes}. 

Flash memories are nonvolatile storage units (i.e., storage units
that remain operational when unpowered), and are usually used for
archival or long-term storage. Information is organized in blocks
of cells, all of which have to be processed jointly during information
erasure cycles. The gist of the approach underlying permutation coding
in flash memories, which uses the fact that the memories consist of
specially organized cells storing charges, is that information is
represented via the relative order of charge levels of cells rather
than their absolute charge levels~\cite{bruck2009rank-modulation}.
This approach, termed \emph{rank modulation}, alleviates the problems
of cell over-injection, reduces the need for block erasures, and is
more robust to errors caused by charge leakage~\cite{bruck2009rank-modulation}.
For instance, while \textit{all} absolute values are subject to errors
caused by charge leakage, the relative ordering of the quantitative
data may remain largely unchanged~\cite{Bruck}. The modeling assumption
behind rank modulation is that only errors swapping adjacently ranked
cell charges are likely~\cite{Bruck,barg2012codes}. As a result,
code design for flash memories was mainly performed in the domain
of the Kendall \textgreek{t} metric, which accounts for small magnitude
errors causing swaps of adjacent elements. A thorough treatment of
codes in the Kendall metric may be found in~\cite{barg2012codes}
and references therein.

In contrast, a more general error model was proposed by the authors
in~\cite{farnoud2013error-correction}, based on the observation
that increasing the number of charge levels in order to increase capacity
decreases the difference between adjacent charge levels and thus unwanted
variations in the charge of a cell may cause its rank to rise above
or fall below the ranks of several other cells instead of only swapping
two adjacent ranks. In addition, the proposed \textit{translocation
error} model adequately accounts for more general types of error such
as read-disturb and write-disturb errors. In this context, the distance
measure of interest is the Ulam distance, related to the length of
the longest common subsequence of two permutations and consequently,
the deletion/insertion or edit distance~\cite{levenshtein_perfect}.
The Ulam distance has also received independent interest in the bioinformatics
and the computer science communities for the purpose of measuring
the ``sortedness'' of data~\cite{Gopalan2007Estimatingsortedness}.
Other metrics used for permutation code construction include the Hamming
distance~\cite{vinckCodedModulation,chu2004constructions} and the
Chebyshev distance (the $\ell_{\infty}$ metric)~\cite{klove2010PermutationArrays,schartz2010correctinglimitedmagnitude}.

Multipermutation codes are a generalization of permutation codes where
each message is encoded as a permutation of the elements of a multiset.
Multipermutation codes in the Hamming metric, known as \emph{constant
composition codes }or \emph{frequency permutation array}s (FPAs),
were studied in several papers including~\cite{Luo2003CCCZq,Ding2005CombinatorialCCC,Huczynska2006FrequencyPermutationArrays,Chu2006OnCCCs}.
For nonvolatile memories, multipermutation coding was proposed by
En Gad et al.~\cite{engad2012trad-offs}, as well as by Shieh and
Tsai~\cite{Shieh2010Decoding}. These works were motivated by different
considerations -- the former aiming to increase the number of possible
re-writes between block erasures, and the latter focusing on the advantages
of multipermutation coding with respect to cell leakage, over-injection
issues, and charge fluctuations. In addition, multipermutation codes
were also recently reported for the Chebyshev distance in~\cite{Shieh2010Decoding,Shieh2011Computingtheball}
and for the Kendall \textgreek{t} distance in~\cite{buzaglo2013error,sala2013dynamic}.

Here, we continue our study of codes in the Ulam metric for nonvolatile
memories by extending it to the level of multipermutation codes. Our
results include bounds on the size of the largest multipermutation
codes, code constructions using multipermutation codes in the Hamming
metric and interleaving as well as permutation codes in the Ulam metric~\cite{farnoud2013error-correction,farnoud2012translocation}.
In the process of analyzing these schemes, we establish new connections
between resolvable balanced incomplete block designs (RBIBDs)~\cite{rumov1976existence,khare1981simple},
semi-Latin squares~\cite{Preece1983SemiLatin}, and multipermutation
codes in the Ulam metric.  As multipermutation codes in the Hamming
metric are used in our constructions, we also provide new bounds on
the size of these codes, and find their asymptotic capacity. In addition,
our results include simple decoding schemes for the proposed constructions
based on designs and those based on interleaving permutation codes
in the Ulam metric.

The paper is organized as follows. In Section~\ref{sec:Preliminaries-and-Notation},
we present the notation used throughout the paper as well as formal
definitions regarding multipermutation codes. In addition, this section
includes motivating examples for our work. Section~\ref{sec:bounds}
is devoted to bounds on the size of multipermutation codes in the
Ulam and Hamming metrics, as well as to the computation of the asymptotic
capacity of these codes. Section~\ref{sec:Constructions} provides
constructions for codes in the Ulam metric. We conclude the paper
in Section~\ref{sec:Conclusion} with a summary of our results and
a number of remarks.

\section{Preliminaries and Notation\label{sec:Preliminaries-and-Notation}}

\begin{figure*}
\captionsetup[subfigure]{labelformat=empty}\subfloat[$\mp=(2,1,1,2),$ $\op=\left(\{2,3\},\{1,4\}\right)$]{\includegraphics[width=0.5\textwidth]{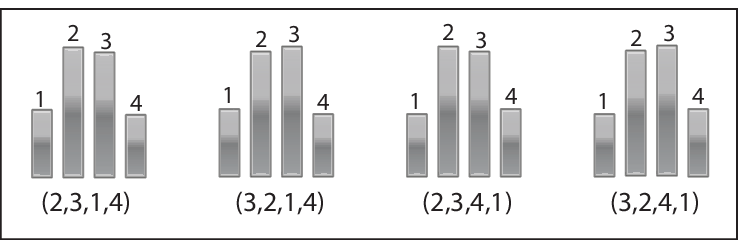}

}\subfloat[$\mp=(1,2,1,2),$ $\op=\left(\{1,3\},\{2,4\}\right)$]{\includegraphics[width=0.5\textwidth]{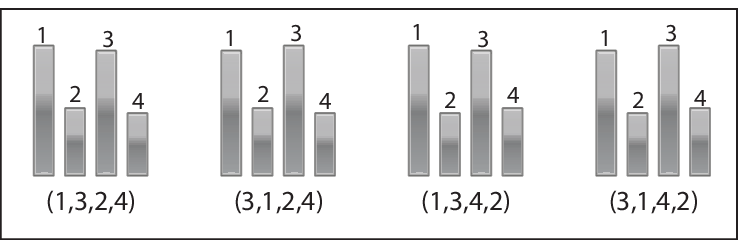}

}

\caption{The two equivalence classes of the code given in~(\ref{eq:ex-mpcode}).
The numbers on the top of the bars indicate cell indices, while the
heights of the bars represent the charge levels of the indicated cells.
The equivalence classes on the left and right correspond to the multipermutations
$\mp=(2,1,1,2)$ and $\mp=(1,2,1,2)$, respectively. Note that each
multipermutation can be programmed into the memory as four different
permutations, each representing a complete ranking of cell charges
without ties.}
\label{fig:mp-code-example}
\end{figure*}

\subsection{Multipermutations, ordered set partitions, and codes}

For an integer $k$, let $[k]=\{1,\dotsc,k\}$. Furthermore, let $\mathbb{S}_{n}$
denote the symmetric group of order $n!$, i.e., the set of permutations
of $n$ distinct elements (typically the elements of $[n]$).

A multipermutation is an arrangement of the elements of a multiset.
For example, $(2,1,2,3,1,2)$ is a multipermutation of $\{1,1,2,2,2,3\}$.
For a positive integer $n$ and a multiplicity vector $\vec{r}=(r_{1},\dotsc,r_{m})$,
such that $n=\sum_{i=1}^{m}r_{i}$, we use $\ms n{\vec{r}}$ to denote
the multiset 
\[
\{\underbrace{1,\dotsc,1}_{r_{1}},\underbrace{2,\dotsc,2}_{r_{2}},\dotsc,\underbrace{m,\dotsc,m}_{r_{m}}\}.
\]
A multiset that has $r$ copies of each of its elements is termed
an \emph{$r$-regular multiset}. For brevity, we henceforth denote
the $r$-regular multiset $\ms n{(r,\cdots,r)}$ by $\ms nr$. An
\emph{$r$-regular multipermutation} is a permutation of an $r$-regular
multiset. Throughout the paper, we focus on $r$-regular multipermutations.
Many of the subsequently described results, however, can easily be
extended to multipermutations of $\ms n{\vec{r}}$ for general multiplicity
vectors $\vec{r}$.

Let $n$ also denote the number of cells in a block of a flash memory.
We assume that $r$ is a positive integer that divides $n$. Consider
a permutation $\pi\in\mathbb{S}_{n}$ that lists the cells in decreasing
order of charge. For example, $\pi=(3,2,4,1)$ means that cell $3$
has the highest charge, cell 2 has the second-highest charge, and
so on. The inverse of $\pi$ is the vector of the ranks of the cells,
$\pi^{-1}=(4,2,1,3)$; cell $1$ has rank $4$, cell $2$ has rank
$2$, and so on. 

To obtain an $r$-regular multipermutation of cell rankings, instead
of assigning rank $i$ to the element in position $i$ in $\pi$,
we assign rank $i$ to all the elements in positions $\{(i-1)r+1,\dotsc,ir\}$.
This multipermutation is denoted by $\mlp r{\pi}$, where 
\[
\mlp r{\pi}(j)=i\mbox{, iff }(i-1)r+1\le\pi^{-1}(j)\le ir.
\]
For instance, given $n=4,r=2$, and $\pi=(3,2,4,1)$, we have $\mlp 2{\pi}=(2,1,1,2)$. 

Observe that for $\pi\in\mathbb{S}_{n}$, $\mlp r{\pi}$ is a multipermutation
of $\ms nr$ and that for $r=1$, the multipermutation $\mlp r{\pi}$
reduces to the inverse of $\pi$, i.e., $\mlp 1{\pi}=\pi^{-1}$. 

A multipermutation $\mp$ of $\ms nr$ can also be represented as
an ordered set partition $\op$, where the $i$th part of $\op$ is
the set 
\[
\op(i)=\left\{ j:\mp(j)=i\right\} .
\]
This definition can naturally be extended to multipermutations of
other multisets. For $\pi\in\mathbb{S}_{n}$, let $\orp r{\pi}$ be
an ordered set partition where 
\[
\orp r{\pi}(i)=\left\{ j:\mlp r{\pi}(j)=i\right\} .
\]
For the aforementioned example $\pi=(3,2,4,1)$, we have $\orp 2{\pi}=(\{2,3\},\{1,4\})$.

For $\pi,\sigma\in\mathbb{S}_{n}$, we write $\pi\eqv r\sigma$ if
$\mlp r{\pi}=\mlp r{\sigma}$, and $\pi\noteqv r\sigma$ otherwise.
It is easy to show that $\eqv r$ is an equivalence relation. The
equivalence class of permutations including $\pi$ is denoted by $\eqclass r(\pi)$,
i.e., 
\[
\eqclass r(\pi)=\{\sigma:\sigma\eqv r\pi\}.
\]

As an illustration, the equivalence class of $(3,2,4,1)$ under $\eqv 2$
equals 
\begin{multline}
\eqclass 2((3,2,4,1))=\\
\{(3,2,4,1),(2,3,4,1),(3,2,1,4),(2,3,1,4)\}.\label{eq:2112}
\end{multline}
 In this case, the set $\eqclass 2((3,2,4,1))$ is isomorphic to the
subgroup $\mathbb{S}_{2}\times\mathbb{S}_{2}$ of $\mathbb{S}_{4}$. 

Let $S$ be a set of size $n$. $ $An \emph{$r$-regular multipermutation
code} $\mpcode{}(n,r)$ over $S$ is a code $C$ whose codewords are
permutations of $S$ with the property that for any $\pi\in C$, $\eqclass r(\pi)\subseteq C$.
For example, 
\begin{multline}
\{(2,3,1,4),(3,2,1,4),(2,3,4,1),(3,2,4,1),\\
(1,3,2,4),(3,1,2,4),(1,3,4,2),(3,1,4,2))\}\label{eq:ex-mpcode}
\end{multline}
is an $\mpcode{}(4,2)$ code. We typically assume that $S=[n]$, but
the results hold for any set $S$ of size $n$.

Each permutation in $C$ represents an ordering of cell charges. For
example $\pi=(1,3,2,4)$ indicates that the cell 1 has the highest
charge, followed by cell 3, and so on. In multipermutation coding,
each $r$ cells are assigned the same rank and all permutations corresponding
to the same multipermutation encode the same information. In the previous
example, the multipermutation $\left(2,1,1,2\right)$ may be represented
by any of the permutations on the right side of (\ref{eq:2112}).

As a result, it is clear that an $\mpcode{}(n,r)$ code $C$ can be
represented as a set of multipermutations $\Mlp rC$, where 
\[
\Mlp rC=\{\mlp r{\pi}:\pi\in C\},
\]
or as a set of ordered set partitions $\Orp rC$, where
\[
\Orp rC=\{\orp r{\pi}:\pi\in C\}.
\]
For example, if $C$ is the code given in~(\ref{eq:ex-mpcode}),
we have 
\begin{align*}
\Mlp 2C & =\{(2,1,1,2),(1,2,1,2))\},\\
\Orp 2C & =\left\{ \left(\{2,3\},\{1,4\}\right),\left(\{1,3\},\{2,4\}\right)\right\} .
\end{align*}
With slight abuse of notation, for an $\mpcode{}(n,r)$ $C$, we may
use $C$ to mean $\Mlp rC$ or $\Orp rC$ if doing so does not lead
to ambiguity. Similarly, we may consider $C$ to be a set of multipermutations
or a set of $ $ordered set partitions instead of a set of permutations.

The \emph{cardinality} or the \emph{size} of $C$, denoted by $|C|$,
equals the number of multipermutations in $\Mlp rC$, or equivalently,
the number of equivalence class of $C$ under the relation $\eqv r$.

In what follows, we describe why multipermutation formats are suitable
for flash memory coding applications. We start with the readback process.
To be able to read the information stored in a flash memory, cells
with different ranks must have charge levels that differ by at least
a certain amount $\Delta$, since if the difference between charge
levels of two cells is too small, it cannot be reliably decided which
one had the higher charge level. Hence, in permutation coding, to
store a permutation of length $n$, the range of possible charge values
must be at least $n\Delta$ to allow for $n$ different charge levels
corresponding to $n$ different ranks. In contrast, an $r$-regular
multipermutation of length $n$ has only $n/r$ ranks and thus it
can be stored in a flash memory whose range of possible charge level
values is $n\Delta/r$. Specifically, the relative order of charge
levels of cells of the same rank of a multipermutation is irrelevant
as all possibilities correspond to the same multipermutation, i.e.,
the same information message.

Note that in order to store information represented by $r$-regular
multipermutations, charges are injected to achieve a desired multipermutation
ranking. As it is neither necessary nor possible for cells of the
same rank to have precisely the same charge levels, the actual representation
of such a multipermutation $\mp$ is some permutation $\pi$, such
that $\mlp r{\pi}=\mp$. The multipermutation is available to the
user retrieving information in the form of the cell charge ordering
$\pi$. As an illustration, consider Figure~\ref{fig:mp-code-example}
for the code given in~(\ref{eq:ex-mpcode}). For instance, to store
the multipermutation $(2,1,1,2)$, any of the permutations given in~(\ref{eq:2112})
may be programmed into the memory. To retrieve the information, the
user reads the permutation, or possibly an erroneous copy of it, and
performs error correction to identify the multipermutation corresponding
to the stored permutation.

Next, we show how multipermutations can achieve a higher information
rate compared to permutations. Consider a flash memory that can accommodate
$m$ sufficiently spread charge levels. In a group of $m$ cells of
such a device, one can store a permutation of length $m$. Suppose
$r$ is a positive integer. It follows that in $mr$ cells, the number
of possible messages that can be stored is $\left(m!\right)^{r}$. 

On the same device, one can store an $r$-regular multipermutation
of length $mr$ in $mr$ cells. In this case, the number of possible
messages equals to the number of possible multipermutations, i.e.,
$\frac{\left(mr\right)!}{\left(r!\right)^{m}}$. It is clear that
for $r\ge2$, we have 
\[
\frac{\left(mr\right)!}{\left(r!\right)^{m}}>\left(m!\right)^{r}.
\]
Hence, in this setting, more information messages can be stored if
one uses multipermutations instead of permutations.

As an illustration, suppose that $m=2$ and $r=10$. Using multipermutations,
we can store 
\[
\lg\frac{\left(mr\right)!}{\left(r!\right)^{m}}\approx17.5\ \mbox{bits}
\]
while using permutations, we can store 
\[
\lg\left(m!\right)^{r}=10\ \mbox{bits}
\]
in 20 cells. Note that here we considered the uncoded regime.

The saving in the number of possible charge levels can also be used
to increase the number of possible re-writes before a block erasure
becomes necessary~\cite{engad2012trad-offs}. As an example, suppose
that $5$ charge levels are available. If one uses permutations of
length $5$, it is only possible to write once before an erasure becomes
necessary, and if one uses permutations of length $3$, it is possible
to write twice before an erasure. Encoding with $r$-regular multipermutations
of length $3r$ also provides the ability to write twice before an
erasure. While both methods, permutation coding and multipermutation
coding, allow for writing twice before an erasure, using multipermutations
leads to a higher information storage rate. For further details on
multipermutation re-write codes, we refer the reader to~\cite{engad2012trad-offs}.

\begin{figure*}
\begin{centering}
\captionsetup[subfigure]{labelformat=parens}\subfloat[$\mlp 2{\sigma}=\mlp 2{\omega}=(3,1,2,1,2,3).$]{\includegraphics[width=0.4\textwidth]{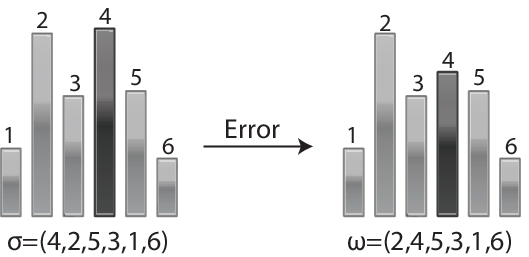}

}\hspace*{2cm}\subfloat[$\mlp 2{\sigma}=(3,1,2,1,2,3),\ \mlp 2{\omega}=(2,1,2,3,1,3).$]{\includegraphics[width=0.4\textwidth]{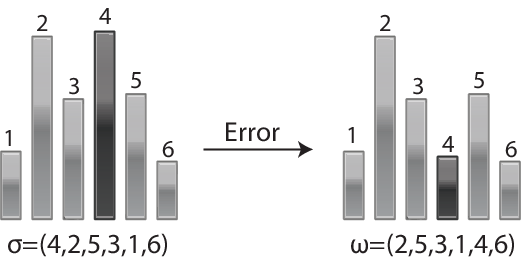}

}
\par\end{centering}

\caption{Examples of errors: A small-magnitude charge drop error (a) manifests
itself as a swap of adjacent ranks in the permutation while a large-magnitude
error (b) manifests itself as a translocation in the permutation.
In multipermutation coding, charge fluctuations may or may not lead
to erroneous multipermutations. Assuming $r=2$, in (a) the multipermutation
does not change, while in (b) it does.}
\label{fig:leakage-example}
\end{figure*}

Before proceeding with an analytical treatment of multipermutation
codes in the Ulam metric, we remark that throughout the paper, we
use $\mathbb{Z}^{+}$ to denote the set of positive integers. Whenever
it is clear from the context, we use the well-known result that $\ln x!=x\ln x+O(x)$,
for any nonnegative real value $x$. By convention, we adopt $0\ln0=0$.

\subsection{The Multipermutation Hamming distance}

For an integer $r$, the \emph{$r$-regular Hamming distance} (or
simply the Hamming distance) $\hdist r$ between two permutations
$\pi,\sigma\in\mathbb{S}_{n}$ is defined as 
\[
\hdist r(\pi,\sigma)=\left|\left\{ i:\mlp r{\pi}(i)\neq\mlp r{\sigma}(i)\right\} \right|.
\]
In words, the permutations are first converted into multipermutations,
which are subsequently compared coordinate-wise. The distance $\hdist r\left(\pi,\sigma\right)$
is equivalent to the ordinary Hamming distance between $\pi$ and
$\sigma$. Thus, instead of $\hdist 1$, we write $\hdist{}$. 

We observe that 
\begin{align*}
\hdist r(\pi,\sigma) & =\sum_{i=1}^{n/r}\left(r-\left|\orp r{\pi}(i)\cap\orp r{\sigma}(i)\right|\right)\\
 & =\sum_{i=1}^{n/r}\left(\left|\orp r{\pi}(i)\right|-\left|\orp r{\pi}(i)\cap\orp r{\sigma}(i)\right|\right)\\
 & =\sum_{i=1}^{n/r}\left|\orp r{\pi}(i)\backslash\orp r{\sigma}(i)\right|.
\end{align*}
Furthermore,
\begin{align*}
\hdist r(\pi,\sigma) & =\min_{\pi'\in R(\pi)}\min_{\sigma'\in R(\sigma)}\hdist{}(\pi',\sigma').
\end{align*}

Let $C$ be an $\mpcode{}(n,r)$ code. The code $C$ has minimum Hamming
distance $d$ if for all $\pi,\sigma\in C$ with $\pi\noteqv r\sigma$,
we have $\hdist r(\pi,\sigma)\ge d$. Equivalently, since for all
$\pi\in C$, $\eqclass r(\pi)$ is contained in $C$, the code $C$
has minimum Hamming distance $d$ if for all $\pi,\sigma\in C$ with
$\pi\noteqv r\sigma$, it holds that $\hdist{}(\pi,\sigma)\ge d$.
An $\mpcode{}(n,r)$ code with minimum Hamming distance $d$ is said
to be an $\mpcode H(n,r,d)$ code. The code given in~(\ref{eq:ex-mpcode})
is an $\mpcode H(4,2,2)$ code.

\subsection{Translocation errors and the Ulam distance}

Figure~\ref{fig:leakage-example} illustrates examples of errors
in flash memories that may occur due to charge leakage, read-disturb,
and write-disturb~\cite{Grupp2009Characterizing}. While errors with
small magnitude represent swaps of adjacent ranks, errors with large
magnitude represent \emph{translocations}~\cite{farnoud2013error-correction}. 

A translocation $\phi(i,j)$ is a permutation that is obtained from
the identity permutation $e$ by moving element $i$ to the position
of $j$ and shifting all elements between $i$ and $j$, including
$j$, by one~\cite{farnoud2013error-correction}. For example, for
$i<j$,
\[
\phi(i,j)=(1,\dotsc,i-1,i+1,i+2,\dotsc,j,i,j+1,\dotsc,n).
\]
As a convention, we assume that $\phi(i,i)=e$. A \emph{translocation
error} is an error that changes a stored permutation $\pi$ to $\pi\phi(i,j)$,
with $i\neq j$. 

A\emph{ subsequence} of a vector $x=\left(x\left(1\right),x\left(2\right),\dotsc,x\left(n\right)\right)$
is a sequence $\left(x\left(i_{1}\right),x\left(i_{2}\right),\dotsc,x\left(i_{k}\right)\right)$,
where $i_{1}<i_{2}<\dotsm<i_{k}$ and $k\le n$. A \emph{common subsequence
}of two vectors $x$ and $y$ is a sequence that is a subsequence
of both $x$ and $y$. Let the length of the longest common subsequence
of two permutations $\pi$ and $\sigma$ be denoted by $\lcs{\pi}{\sigma}$.
The \emph{Ulam distance $\tdist{}\left(\pi,\sigma\right)$} between
two permutations $\pi$ and $\sigma$ of length $n$ is defined as
$ $$n-\lcs{\pi}{\sigma}$. It is straightforward to see that the
Ulam distance between $\pi$ and $\sigma$ equals the minimum number
of translocations required to take $\pi$ to $\sigma$~\cite{farnoud2013error-correction}.
It is also well known that the Ulam distance represents the \textit{edit
distance} between two permutations, i.e., the smallest number of insertion/deletion
pairs needed to transform one permutation into another.

For $\pi,\sigma\in\mathbb{S}_{n}$, define the \emph{($r$-regular)
Ulam distance} $\tdist r$ on permutations as
\[
\tdist r(\pi,\sigma)=\min_{\pi'\in R(\pi)}\min_{\sigma'\in R(\sigma)}\tdist{}(\pi',\sigma').
\]
Note that this distance is a set-distance: it measures the smallest
Ulam distance between two permutations in different equivalence classes.
Furthermore, the distance $\tdist r(\pi,\sigma)$ equals the minimum
number of translocations required to take a permutation in $\eqclass r(\pi)$
to a permutation in $\eqclass r(\sigma)$. 

Let $\closestpair r{\pi}{\sigma}$ denote the set 
\[
\{(\alpha,\beta):\alpha\in\eqclass r(\pi),\beta\in\eqclass r(\sigma),\tdist r(\pi,\sigma)=\tdist{}(\alpha,\beta)\}.
\]
By definition of $\tdist r(\pi,\sigma)$, $\closestpair r{\pi}{\sigma}$
is nonempty.

An $\mpcode{}(n,r)$ $C$ has minimum Ulam distance $d$ if for all
$\pi,\sigma\in C$ with $\pi\noteqv r\sigma$, we have $\tdist r(\pi,\sigma)\ge d$.
Such a code is denoted by $\mpcode{\circ}(n,r,d)$. The code given
in~(\ref{eq:ex-mpcode}) is an $\mpcode{\circ}(4,2,1)$ code, as
$\tdist{}\left((3,2,1,4),(3,1,2,4)\right)=1.$

Under minimum distance decoding, an $\mpcode{\circ}(n,r,d)$ code
can correct $t$ translocation errors iff $d\ge2t+1$. To see this,
note that $d<2t+1$ iff there exists $\pi,\sigma\in C,\pi\noteqv r\sigma,$
and $\omega\in\mathbb{S}_{n}$ such that $\tdist{}(\omega,\pi)\le t$
and $\tdist{}(\omega,\sigma)\le t$, iff $C$ cannot correct $t$
errors.

For a set $P$ and a permutation $\pi$, let $\pi_{P}$ denote the
projection of $\pi$ onto $P$, that is, the sequence obtained by
only keeping those elements of $\pi$ that are in $P$. We find the
following lemma, proved in our companion paper~\cite{farnoud2013error-correction},
useful in our subsequent derivations. 
\begin{lem}
\label{lem:projection}For sets $P\subseteq[n]$ and $Q=[n]\backslash P$,
and for permutations $\pi,\sigma\in\mathbb{S}_{n}$, we have 
\[
\tdist{}(\pi,\sigma)\ge\tdist{}(\proj{\pi}P,\proj{\sigma}P)+\tdist{}(\proj{\pi}Q,\proj{\sigma}Q).
\]

\end{lem}

\subsection{Relationship between the Ulam and the Hamming metrics}

The following lemma is an immediate consequence of the definition
of a translocation.
\begin{lem}
\label{lem:Atranslocation}A translocation, applied to a permutation,
changes at most one element of each rank. That is, for a translocation
$\varphi$, a permutation $\pi$, and $i\in[n/r]$, 
\[
\left|\orp r{\pi}(i)\cap\orp r{\pi\varphi}(i)\right|\ge r-1.
\]

\end{lem}
Since there are $n/r$ ranks, we have
\[
\hdist r(\pi,\pi\varphi)\le\frac{n}{r},
\]
for a translocation $\varphi$ and a permutation $\pi$. Hence, 
\[
\hdist r(\pi,\sigma)\le\frac{n}{r}\tdist r(\pi,\sigma)
\]
for $\pi,\sigma\in\mathbb{S}_{n}$.

We next upper bound $\tdist r(\pi,\sigma)$ in terms of $\hdist r(\pi,\sigma).$
There exist $\pi'\in\eqclass r(\pi)$, $\sigma'\in\eqclass r(\sigma)$,
and a common subsequence of $\pi'$ and $\sigma'$ that contains the
elements of 
\[
\bigcup_{i=1}^{n/r}\left(\orp r{\pi}(i)\cap\orp r{\sigma}(i)\right).
\]
Hence,
\begin{align*}
\tdist r(\pi,\sigma) & \le n-\lcs{\pi'}{\sigma'}\\
 & \le n-\sum_{i=1}^{n/r}\left|\orp r{\pi}(i)\cap\orp r{\sigma}(i)\right|\\
 & =\sum_{i=1}^{n/r}\left(r-\left|\orp r{\pi}(i)\cap\orp r{\sigma}(i)\right|\right)\\
 & =\hdist r(\pi,\sigma),
\end{align*}
implying that $\tdist r(\pi,\sigma)\le\hdist r(\pi,\sigma).$ 
\begin{lem}
\label{lem:HammingUlam}For $\pi,\sigma\in\mathbb{S}_{n}$, we have
\[
\frac{r}{n}\hdist r(\pi,\sigma)\le\tdist r(\pi,\sigma)\le\hdist r(\pi,\sigma).
\]

\end{lem}
The lemma illustrates the fact that for $r=\Theta(n)$, the Ulam distance
is within a constant factor of the Hamming distance, while for $r=o(n)$,
the Ulam distance may be much smaller. Consequently, while good codes
in the Ulam metric allow for substitution error correction, good codes
in the Hamming metric provide resilience under translocation errors
only for a certain limited range of parameters.

\section{Bounds on Size of Multipermutation Codes\label{sec:bounds}}

In what follows, we derive bounds on the size of multipermutation
codes in the Hamming metric as well as the Ulam metric. For the case
of the Hamming distance, we find the asymptotic capacity, while for
the Ulam distance we provide lower and upper bounds on the capacity.
We point out that a number of bounds on multipermutation codes in
the Hamming metric were derived in~\cite{Huczynska2006FrequencyPermutationArrays},
including some simple and some complicated expressions involving Laguerre
polynomials. Nevertheless, these bounds do not allow for finding a
capacity formula for the underlying codes.

Let $A_{H}(n,r,d)$ and $A_{\circ}(n,r,d)$ denote the maximum cardinalities
of an $\mpcode H(n,r,d)$ and an $\mpcode{\circ}(n,r,d)$ code, respectively.
Furthermore, let $\mathcal{C}_{H}(r,d)$ denote the \emph{capacity},
i.e., maximum achievable rate, of multipermutation codes in the Hamming
metric, defined as 
\[
\mathcal{C}_{H}(r,d)=\lim_{n\to\infty}\frac{\ln A_{H}(n,r,d)}{\ln n!}\cdot
\]
The capacity of multipermutation codes in the Ulam metric, $\mathcal{C}_{\circ}(r,d)$,
is defined similarly.

In the remainder of the paper, limits are evaluated for $n\to\infty$
unless stated otherwise. We assume that all limits of interest exist
and we use $\rho=\rho(r)=\lim\frac{\ln r}{\ln n}$, as well as $\delta=\delta(d)=\lim\frac{d}{n}\cdot$

\subsection{Multipermutation Codes in the Hamming Metric }

It was shown by Luo et al.~\cite{Luo2003CCCZq} that
\begin{equation}
A_{H}(n,r,d)\le\frac{d}{r+d-n},\quad\mbox{for }r+d>n,\label{eq:luo-UB}
\end{equation}
 and by Huczynska and Mullen~\cite{Huczynska2006FrequencyPermutationArrays}
that
\begin{equation}
A_{H}(n,r,d)\le\frac{n!}{r(d-1)!}\cdot\label{eq:Huc-UB}
\end{equation}
The first bound,~(\ref{eq:luo-UB}), implies that the asymptotic
rate is zero if $r+d>n$, while the second bound implies that 
\begin{equation}
\mathcal{C}_{H}(r,d)\le1-\delta,\label{eq:simpleUB}
\end{equation}
 which also follows from the fact that $\mathcal{C}_{H}(r,d)\le\mathcal{C}_{H}(1,d)$
and Theorem 11 of our companion paper~\cite{farnoud2013error-correction},
stating that $\mathcal{C}_{H}(1,d)=1-\delta$. We improve next upon
the bound in~(\ref{eq:simpleUB}) and provide a matching lower bound,
thereby establishing the capacity of $r$-regular multipermutation
codes in the Hamming metric.

Let $S(l,m,r)$ denote the number of sequences of length $l$ over
the alphabet $[m]$, with no element appearing more than $r$ times.
Note that $S(l,m,r)$ equals the number of ordered partitions of a
set of size $l$ into $m$ sets such that each part has at most $r$
elements, and where empty subsets are allowed.
\begin{lem}
\label{lem:Hamming-Singleton}(Singleton bound) For positive integers
$n,r,d$ such that $r$ divides $n$, we have 
\[
A_{H}(n,r,d)\le\left(\frac{n}{r}\right)^{n-d+1}.
\]
\end{lem}
\begin{IEEEproof}
Consider an $\mpcode H(n,r,d)$ code $C$ of size $M$ and let 
\[
\Mp=\Mlp rC=\{\mp_{1},\dotsc,\mp_{M}\}
\]
denote its multipermutation representation. Since the minimum Hamming
distance $\hdist r$ of $C$ is at least $d$, for distinct $i$ and
$j$, 
\begin{equation}
\sum_{k=1}^{n}\mathbb{I}\left(\mp_{i}(k)\neq\mp_{j}(k)\right)\ge d,\label{eq:nExcluded-1-1}
\end{equation}
where the indicator function $\mathbb{I}$ is defined in the standard
manner as 
\[
\mathbb{I}(\mathsf{condition})=\begin{cases}
1, & \quad\mbox{if }\mathsf{condition}\mbox{ is true},\\
0, & \quad\mbox{if }\mathsf{condition}\mbox{ is false.}
\end{cases}
\]
By removing the last $d-1$ elements of each multipermutation $\mp_{i},i\in[M],$
we obtain the set $\Mp'=\left\{ \mp'_{1},\dotsc,\mp'_{M}\right\} $
of sequences of length $n-d+1$ over $[n/r]$ where no element appears
more than $r$ times.%
\footnote{This argument is akin to the approach proposed in~\cite{barg2012codes}
for permutation codes in the Kendall metric.%
}

Since $d-1$ elements are removed,~(\ref{eq:nExcluded-1-1}) implies
that 
\[
\sum_{k=1}^{n-d+1}\mathbb{I}\left(\mp'_{i}(k)\neq\mp'_{j}(k)\right)=\sum_{k=1}^{n-d+1}\mathbb{I}\left(\mp{}_{i}(k)\neq\mp{}_{j}(k)\right)\ge1.
\]
and thus for distinct $i,j\in[M]$, $\mp'_{i}$ and $\mp'_{j}$ are
distinct. Hence, we have 
\begin{align}
A_{H}(n,r,d) & \le S\left(n-d+1,\frac{n}{r},r\right).\label{eq:AH_bound_S}
\end{align}
Furthermore, since $S(n-d+1,n/r,r)\le S(n-d+1,n/r,\infty),$ we find
\[
A_{H}(n,r,d)\le S\left(n-d+1,\frac{n}{r},\infty\right)=\left(\frac{n}{r}\right)^{n-d+1}.
\]

\end{IEEEproof}
As shown in the sequel, the bound given in Lemma~\ref{lem:Hamming-Singleton}
is sufficiently tight for capacity derivations. Nevertheless, it may
be useful to bound $S(l,m,r)$ more tightly. 

\begin{table*}
\centering{}\caption{Bounds on the size $3$-regular multipermutation codes in the Hamming
metric of length $9$.}
\label{tab:HammingUBs}%
\begin{tabular}{|c|c|c|c|c|c|c|c|c|c|}
\hline 
Upper bound on $A_{H}(9,3,d)$ & $d=1$ & $d=2$ & $d=3$ & $d=4$ & $d=5$ & $d=6$ & $d=7$ & $d=8$ & $d=9$\tabularnewline
\hline 
\hline 
(\ref{eq:luo-UB}) \cite{Luo2003CCCZq} & - & - & - & - & - & - & 7 & 4 & 3\tabularnewline
\hline 
(\ref{eq:Huc-UB}) \cite{Huczynska2006FrequencyPermutationArrays} & 120960 & 120960 & 60480 & 20160 & 5040 & 1008 & 168 & 24 & 3\tabularnewline
\hline 
Lemma \ref{lem:Hamming-Singleton} & 19683 & 6561 & 2187 & 729 & 243 & 81 & 27 & 9 & 3\tabularnewline
\hline 
(\ref{eq:CLT-bonud}) (approximate bound) & 12077 & 4560 & 1700 & 624 & 224 & 79 & 27 & 9 & 3\tabularnewline
\hline 
(\ref{eq:EGF}) & 1680 & 1680 & 1050 & 510 & 210 & 78 & 27 & 9 & 3\tabularnewline
\hline 
\end{tabular}
\end{table*}

It is easy to see that 
\begin{equation}
S(l,m,r)=\sum_{\begin{array}{c}
x_{1}+\cdots+x_{m}=l,\\
0\le x_{i}\le r
\end{array}}\frac{l!}{\prod_{i=1}^{m}x_{i}!}\cdot\label{eq:mulinom}
\end{equation}
where the $x_{i}$'s are integers. The exponential generating function
(EGF) of $S(l,m,r)$ is
\[
\sum_{l=0}^{\infty}S(l,m,r)\frac{z^{l}}{l!}=\left(\sum_{i=0}^{r}\frac{z^{i}}{i!}\right)^{m}
\]
and thus one can write 
\begin{equation}
A_{H}(n,r,d)\le(n-d+1)!\left[z^{n-d+1}\right]\left(\sum_{i=0}^{r}\frac{z^{i}}{i!}\right)^{n/r}\cdot\label{eq:EGF}
\end{equation}
The bound given in~(\ref{eq:EGF}) can be used to find numerical
upper bounds on the code size, such as those provided in Table~\ref{tab:HammingUBs}.
In addition, it can be used to obtain simple asymptotic bounds using
methods described in the classical text~\cite[Ch.~8]{odlyzko_asymptotic_1995}.
As a final note, we point out that the related problems of restricted
multisets and restricted integer partitions are far better studied
combinatorial entities than the one we addressed above~\cite[Ch.~21, Sec.~8]{Graham1995Handbook},
although no simple direct connection between these problems and the
problem discussed here exist.

Another approach, which is conceptually much simpler and which applies
to many other coding-theoretic scenarios is using the Poisson approximation
theorem for multinomial variables and the Chernoff bound~\cite{Deheuvels1988Poissonapproximations,Arenbaev1976AsymptoticBehavioroftheMultinomialDstribution},
or alternatively, the Central limit theorem~\cite{hofri1995analysis}.
As shown in~\cite{milenkovic2004probUrmModels,hofri1995analysis},
the number of terms in the multinomial summation formula, $m$, may
represent the number of labeled urns into which $l$ labeled balls
are thrown randomly. The occupancy variables $X_{i}$, $i=1,\ldots,m$,
are dependent, since $X_{1}+\ldots+X_{m}=l$. But in the asymptotic
central domain regime, with $l/m$ constant, the variables $X_{i}$,
$i=1,\ldots,m$, may be viewed as \emph{independent} Poisson variables
with mean $\lambda=l/m$. Any result of computations involving independent
Poisson variables that satisfies the inversion conditions dictated
by Tauberian theorems described in~\cite{milenkovic2004probUrmModels}
may be \emph{asymptotically converted} into the correct result by
simply replacing $\lambda$ with $l/m$. Furthermore, the same approach
may be used when dealing with urns and balls that satisfy additional
constraints~\cite{milenkovic2004probUrmModels}.

To understand the principles behind the Poisson transform method,
we follow the analysis in~\cite{hofri1995analysis} based on\ \cite{gonnet1984analysis}.
The key observation is that the Poisson distributions satisfy the
additivity (infinite divisibility) property, i.e., the property that
the sum of independent Poisson random variables is another Poisson
random variable with mean parameter equal to the sum of the parameters
of the individual variables in the sum. Then, it is straightforward
to show that for two different ball placement processes, the urn occupancy
variables have the same distribution: 1) in the first case, each urn
receives balls according to a Poisson distribution with parameter
$\lambda$ independently of all other urns; 2) in the second case,
balls arrive with a Poisson distribution with parameter $\lambda m$
and are routed with uniform probability $1/m$ to one of the urns. 

Assume next that $g(m,\lambda)$ is a quantity of interest where the
input to each urn is generated according to model 1). The same quantity
under the original urns and balls model with a fixed number $l$ of
balls is denoted by $f(m,l)$. Using the equivalence between the two
formulations 1) and 2), one can show that
\[
g(m,\lambda)=\sum_{l}f(m,l)\: P\left(X_{1}+\ldots+X_{m}=l\right),
\]
where $X_{1},\ldots,X_{m}$ are i.i.d Poisson random variables with
parameter $\lambda$. As a result, it is straightforward to see that
$f(m,l)=\frac{l!}{m^{l}}\,\,[\lambda^{l}]\left\{ e^{\lambda\, m}g(m,\lambda)\right\} .$ 

In words, $e^{\lambda\, m}g(m,\lambda)$ represents the exponential
generating function over the number of balls $l$ of $f(m,l)$ evaluated
at $\lambda\, m$. Evaluating the coefficient in a generating function
in the asymptotic domain may be accomplished with the aid of Tauberian
theorems (see~\cite{milenkovic2004probUrmModels}) or classical asymptotic
analysis. In the case of the Poisson transform, provided that some
minor technical conditions are met, it can be shown that $f(m,l)\simeq\, g(m,l/m),$
where $a(x)\simeq b(x)$ stands for $\lim_{x\to\,\infty}a(x)/b(x)=1$.
Intuitively, the aforementioned result implies that when the dependencies
among a large number of random variables are weak -- for example,
only in terms of a constraint on the total sum of their values --
then the variables are asymptotically independent, provided a proper
choice of the distribution ensures consistence with the finite-valued
parameters.

In the case of interest, we need to find the probability $P\{X_{i}\leq r,\, i=1,\ldots,m\}$.
For $m,l\to\infty$, such that $l/m$ is a constant, and for $r$
fixed, this leads to
\begin{equation}
S(l,m,r)\simeq m^{l}\,\left(\sum_{i=0}^{r}\,\exp(-\lambda)\,\frac{\lambda^{i}}{i!}\right)^{m},\notag
\end{equation}
where $\lambda=l/m$. The asymptotic formula for $S(l,m,r)$ depends
on the relationship between the parameters $r,l,m.$ For $r\leq l/m$,
the Chernoff bound reads as
\[
\sum_{i=0}^{r}\,\exp(-\lambda)\,\frac{\lambda^{i}}{i!}\leq\frac{\exp(-\lambda)(e\,\lambda)^{r}}{r^{r}},
\]
so that
\begin{equation}
S(l,m,r)\lesssim m^{l}\,\frac{\exp(-l/m)(e\, l/m)^{r}}{r^{r}}\cdot
\end{equation}

For the case of interest in our derivation, $r>l/m$. Whenever $r>10$,
one may use the straightforward Central Limit Theorem approximation
\[
\sum_{i=0}^{r}\,\exp(-\lambda)\,\frac{\lambda^{i}}{i!}\simeq\Phi\left(\frac{r+0.5-l/m}{\sqrt{l/m}}\right),
\]
so that
\[
S(l,m,r)\simeq m^{l}\;\Phi^{m}\left(\frac{r+0.5-l/m}{\sqrt{l/m}}\right),
\]
where the function $\Phi(\cdot)$ stands for the cumulative distribution
function (CDF) of a standard Gaussian random variable. Since $r>\frac{n-d+1}{n/r}$,
the preceding relation and (\ref{eq:AH_bound_S}) imply
\begin{equation}
A_{H}(n,r,d)\lesssim\left(\frac{n}{r}\right)^{n-d+1}\;\Phi^{n/r}\left(\frac{n+2(d-1)r}{2\sqrt{n(n-d+1)r}}\right)\label{eq:CLT-bonud}
\end{equation}
 provided that $\frac{n-d+1}{n/r}$ is a constant larger than 10.

As an example, upper bounds on the size of $\mpcode H(9,3,d)$ are
given in Table~\ref{tab:HammingUBs}. Note that the bounds of~(\ref{eq:CLT-bonud}),
Lemma~\ref{lem:Hamming-Singleton}, and~(\ref{eq:EGF}) are very
close for small values of $d$. Indeed, the right side of~(\ref{eq:CLT-bonud})
is bounded above by $(n/r)^{n-d+1}$ and below by 
\[
\left(\frac{n}{r}\right)^{n-d+1}\left(\frac{1}{2}\right)^{n/r}
\]
and we have 
\[
\lim\frac{\ln\left(\left(\frac{n}{r}\right)^{n-d+1}\left(\frac{1}{2}\right)^{n/r}\right)}{\ln\left(\frac{n}{r}\right)^{n-d+1}}=1
\]
provided that $\delta<1$ and $r<n$. Therefore, the bounds of~(\ref{eq:CLT-bonud})
and Lemma \ref{lem:Hamming-Singleton} have the same asymptotic exponent.

The next lemma provides a lower bound on $A_{H}(n,r,d)$.
\begin{lem}
\label{lem:Hamming-Gilbert-Varshamov}(Gilbert-Varshamov Bound) We
have 
\[
A_{H}(n,r,d)\ge\frac{n!}{(r!)^{n/r}\binom{n}{d-1}\left(\frac{n}{r}\right)^{d-1}}\cdot
\]
\end{lem}
\begin{IEEEproof}
There are $\frac{n!}{(r!)^{n/r}}$ multipermutations of $\ms nr$.
The size of a ball of radius $d-1$ in the space of multipermutations
of $\ms nr$ endowed with the Hamming distance is bounded above by
$\binom{n}{d-1}\left(\frac{n}{r}\right)^{d-1}$ (an exact and complicated
expression for the size of the ball may be found in~\cite{Huczynska2006FrequencyPermutationArrays}).
The Lemma follows by a standard application of Gilbert's argument.\end{IEEEproof}
\begin{thm}
\label{thm:Hamming-cap}We have
\[
\mathcal{C}_{H}(r,d)=(1-\rho)(1-\delta).
\]
\end{thm}
\begin{IEEEproof}
First, recall that $\lim$ expressions with no subscripts stand for
$\lim_{n\to\infty}$. 

On the one hand, from Lemma~\ref{lem:Hamming-Singleton}, we have
\begin{align*}
\mathcal{C}_{H}(r,d) & \le\lim\frac{(n-d+1)(\ln n-\ln r)}{\ln n!}\\
 & =\lim\frac{n\ln n-n\ln r-d\ln n+d\ln r}{n\ln n+O(n)}\\
 & =1-\rho-\delta+\rho\delta.
\end{align*}

On the other hand, from Lemma~\ref{lem:Hamming-Gilbert-Varshamov},
we easily see that
\begin{align*}
\mathcal{C}_{H}(r,d) & \ge\lim\frac{\ln\left(n!(r!)^{-n/r}\binom{n}{d-1}^{-1}\left(\frac{n}{r}\right)^{-d+1}\right)}{\ln n!}\\
 & =1-\lim\frac{(n/r)\ln r!+(d-1)\ln(n/r)}{\ln n!}\\
 & =1-\lim\frac{n\ln r+d\ln n-d\ln r+O(n)}{n\ln n+O(n)}\\
 & =1-\rho-\delta+\rho\delta,
\end{align*}
where we have used the fact that $\lim\frac{\ln\binom{n}{d-1}}{\ln n!}=0$.
This establishes the claimed result for the asymptotic capacity of
multipermutation codes in the Hamming metric.
\end{IEEEproof}

\subsection{Multipermutation Codes in the Ulam Metric}

Using Lemma~\ref{lem:HammingUlam} which implies that $A_{\circ}(n,r,d)\le A_{H}(n,r,d)$,
we find the following upper bound on $A_{\circ}(n,r,d)$:
\begin{equation}
A_{\circ}(n,r,d)\le A_{H}(n,r,d)\le S(n-d+1,\frac{n}{r},r)\le\left(\frac{n}{r}\right)^{n-d+1}.\label{eq:Ulam-maxsize-UB}
\end{equation}
The next lemma provides a lower bound on $A_{\circ}(n,r,d)$.
\begin{lem}
\label{lem:GVB-Ulam}(Gilbert-Varshamov Bound) For positive integers
$n,r,d$ such that $n$ is a multiple of $r$, we have 
\[
A_{\circ}(n,r,d)\ge\frac{(n-d+1)!}{\binom{n}{d-1}(r!)^{2n/r}}\cdot
\]
\end{lem}
\begin{IEEEproof}
Let $\ball r{\circ}(u)$ denote the size of a ball of radius $u$
in $\mathbb{S}_{n}$ endowed by $\tdist r$ (note that due to symmetry,
i.e., left invariance of the Ulam metric, the volume of the ball is
independent on the choice of the center). Equivalently, let $\ball r{\circ}(u)=\left\{ \pi\in\mathbb{S}_{n}:\tdist r(\pi,e)\le u\right\} $.
The Gilbert bound states that 
\[
A_{\circ}(n,r,d)\ge\frac{n!}{\ball r{\circ}(d-1)}\cdot
\]
We show that $\ball r{\circ}(d-1)\le(r!)^{2n/r}\ball 1{\circ}(d-1)$.
The lemma then follows from a result pertaining to the Ulam metric
we derived in~\cite{farnoud2013error-correction}, namely: 
\[
\ball 1{\circ}(d-1)\le\binom{n}{d-1}\frac{n!}{(n-d+1)!}\cdot
\]

The set $\left\{ \pi\in\mathbb{S}_{n}:\tdist r(\pi,e)\le u\right\} $
equals 
\[
\bigcup_{\sigma\in\eqclass r(e)}\ \bigcup_{\pi\in\mathbb{S}_{n}:\tdist{}(\pi,\sigma)\le u}\eqclass r(\pi).
\]
 Hence,
\begin{align*}
\ball r{\circ}(u) & =\left|\cup_{\sigma\in\eqclass r(e)}\cup_{\pi\in\mathbb{S}_{n}:\tdist{}(\pi,\sigma)\le u}\eqclass r(\pi)\right|\\
 & \le(r!)^{n/r}\left|\cup_{\pi\in\mathbb{S}_{n}:\tdist{}(\pi,e)\le u}\eqclass r(\pi)\right|\\
 & \le(r!)^{n/r}\ball 1{\circ}(u)(r!)^{n/r}\\
 & =(r!)^{2n/r}\ball 1{\circ}(u),
\end{align*}
which shows that $ $$\ball r{\circ}(d-1)\le(r!)^{2n/r}\ball 1{\circ}(d-1)$.
\end{IEEEproof}
Next, we improve upon Lemma~\ref{lem:GVB-Ulam} by finding a sharper
bound for $\ball r{\circ}(u),u\in\mathbb{Z}^{+}$. Consider the ball
around the identity permutation $e$. For a permutation $\pi$ that
satisfies $\tdist r(\pi,e)\le u$, there exists a $\pi'\in\eqclass r(\pi)$
that has a common subsequence $s$ of length $l=n-u$ with some $e'\in\eqclass r(e)$.
There are 
\[
A=\sum_{\begin{array}{c}
x_{1}+\cdots+x_{n/r}=l,\\
x_{i}\in[0,r],\forall i
\end{array}}\prod_{i=1}^{n/r}\binom{r}{x_{i}}x_{i}!
\]
 ways of choosing a sequence $s$ of length $l$ such that it is a
subsequence of some $e'\in\eqclass r(e)$, with $\prod_{i=1}^{n/r}\binom{r}{x_{i}}x_{i}!$
counting the number of ways one can choose a subsequence of length
$l$ with $x_{i}$ elements from rank $i$, 
\[
\orp r{e'}(i)=\orp re(i)=\{(i-1)r+1,\dotsc,ir\}.
\]

The number of ordered partitions $\op$ with parts of size equal to
$r$, such that there exists a $\pi'$ that satisfies $\op=\orp r{\pi'}$
and contains $s$ as a subsequence equals
\[
B=\sum_{\begin{array}{c}
x_{1}+\cdots+x_{n/r}=l,\\
x_{i}\in[0,r],\forall i
\end{array}}\binom{n-l}{r-x_{1},\dotsc,r-x_{n/r}}.
\]
Here, the multinomial $\binom{n-l}{r-x_{1},\dotsc,r-x_{n/r}}$ accounts
for the number of ways of choosing the ordered partition $\op$ such
that the first $x_{1}$ elements of $s$ are in the first part, the
next $x_{2}$ elements are in the second part, and so on.

In addition, there are $(r!)^{n/r}$ permutations $\pi$ such that
$\pi\in\eqclass r(\pi')$. Hence, 
\begin{equation}
\ball r{\circ}(u)\le AB(r!)^{n/r}.\label{eq:sizeofBall-1}
\end{equation}

With regards to bounding the combinatorial sum $A$, we observe that
\begin{align*}
A & \le\binom{l+n/r-1}{n/r-1}\max_{\begin{array}{c}
x_{1}+\cdots+x_{n/r}=l,\\
x_{i}\in[0,r],\forall i
\end{array}}\prod_{i=1}^{n/r}\binom{r}{x_{i}}x_{i}!\\
 & \le\binom{l+n/r-1}{n/r-1}\Bigl(\frac{r!}{(r-\frac{l}{n/r})!}\Bigr)^{n/r}\\
 & \le\binom{2n}{n}\Bigl(\frac{r!}{\frac{r(n-l)}{n}!}\Bigr)^{n/r}.
\end{align*}
Using a similar approach for $B$, we find
\[
B\le\binom{2n}{n}\frac{(n-l)!}{\left(\frac{r(n-l)}{n}!\right)^{n/r}}\cdot
\]
 Hence, 
\[
\ball r{\circ}(u)\le\binom{2n}{n}^{2}\Bigl(\frac{r!}{\frac{ru}{n}!}\Bigr)^{2n/r}u!,
\]
 and so, for $d>1$,
\begin{align*}
\ln\ball r{\circ}(d-1) & \le\frac{2n}{r}\ln r!-\frac{2n}{r}\ln\frac{r(d-1)}{n}!\\
 & \qquad+\ln(d-1)!+O(n)\\
 & =2n\ln r-2(d-1)\ln\frac{r(d-1)}{n}\\
 & \qquad+(d-1)\ln(d-1)+O(n)\\
 & =2n\ln r-2d\ln r+2d\ln n-d\ln d+O(n).
\end{align*}
This implies that
\begin{align}
\mathcal{C}_{\circ}(r,d) & \ge1-\lim\frac{2n\ln r-2d\ln r+2d\ln n-d\ln d+O(n)}{n\ln n+O(n)}\nonumber \\
 & =1-2\rho+2\delta\rho-\delta\nonumber \\
 & =(1-\delta)(1-2\rho).\label{eq:Ulam-cap-LB}
\end{align}

\begin{thm}
\label{thm:capacity-Ulam}The capacity of multipermutation codes in
the Ulam metric is bounded according to
\[
(1-\delta)(1-2\rho)\le\mathcal{C}_{\circ}(r,d)\le(1-\delta)(1-\rho).
\]
\end{thm}
\begin{IEEEproof}
The lower bound is given in~(\ref{eq:Ulam-cap-LB}) while the upper
bound is a result of~(\ref{eq:Ulam-maxsize-UB}) and Theorem~\ref{thm:Hamming-cap}.
\end{IEEEproof}

\section{Constructions\label{sec:Constructions}}

In the next subsections, we present several constructions for multipermutation
codes in the Ulam metric. One of the key ingredients of our constructions
is permutation interleaving, which we proposed for Ulam metric code
design in~\cite{farnoud2013error-correction}. The related idea of
restricting certain positions in the codewords to certain values was
first described in~\cite{klove2010PermutationArrays,schartz2010correctinglimitedmagnitude},
while interleaving in the Chebyshev metric was discussed in\cite{Shieh2010Decoding}. 

For sequences $\pi_{1},\dotsc,\pi_{k}$, let $\pi_{1}\circ_{r}\pi_{2}\circ_{r}\dotsm\circ_{r}\pi_{k}$
denote the sequence obtained by sequentially interleaving blocks of
$r$ elements of $\pi_{i},i\in[k]$. 

For example, $(1,3,4,2)\circ_{2}(6,7,8,5)\circ_{2}(12,10,9,11)=(1,3,6,7,12,10,4,2,8,5,9,11)$. 

This form of interleaving will henceforth be called block interleaving.
Whenever $r=1$, we simply write $\circ$ instead of $\circ_{1}$.

\subsection{Constructions based on almost disjoint sets}

Two sets $A$ and $B$ are said to be \emph{at most k-intersecting,
}if for a given positive integer $k$, one has
\[
|A\cap B|\le k.
\]
When $k$ is smaller than the size of the sets $A,B$, and the aforementioned
bound is true, we say that the sets are \textit{almost disjoint}.
The next lemma shows how sets of set partitions with almost disjoint
parts can be used for constructing multipermutation codes in the Ulam
metric.
\begin{lem}
\label{lem:construct-1}Let $C$ be an $\mpcode{}(n,r)$ code, and
suppose that $t$ is a positive integer such that $2t<r$. If for
all $\pi,\sigma\in C$ and $i\in[n/r]$, we either have $\orp r{\pi}(i)=\orp r{\sigma}(i)$
or 
\begin{equation}
\left|\orp r{\pi}(i)\cap\orp r{\sigma}(i)\right|<r-2t,\label{eq:construct-1}
\end{equation}
then the code $C$ can correct $t$ translocation errors, that is,
$C$ is an $\mpcode{\circ}(n,r,2t+1)$ code.\end{lem}
\begin{IEEEproof}
Suppose $\pi\in C$ is the (unknown) stored codeword and $\omega$
is the retrieved permutation. The Ulam distance between $\pi$ and
$\omega$ is at most $t$ since the codeword $\pi$ is affected by
at most $t$ translocation errors. We show that given $\omega$, $\orp r{\pi}$
can be uniquely identified. Fix $i\in[n/r]$. Since there are at most
$t$ translocation errors, by Lemma~\ref{lem:Atranslocation}, we
have 
\begin{equation}
\left|\orp r{\pi}(i)\cap\orp r{\omega}(i)\right|\ge r-t.\label{eq:t-errors}
\end{equation}
To identify $\orp r{\pi}(i)$ uniquely, it suffices to have $\left|\orp r{\sigma}(i)\cap\orp r{\omega}(i)\right|<r-t$
for all $\sigma\in C$ such that $\op_{\pi}^{r}(i)\neq\op_{\sigma}^{r}(i)$. 

Suppose that $\sigma\in C$ and $\op_{\pi}^{r}(i)\neq\op_{\sigma}^{r}(i)$.
We use~(\ref{eq:construct-1}) and~(\ref{eq:t-errors}) to show
that $\left|\orp r{\sigma}(i)\cap\orp r{\omega}(i)\right|<r-t$. For
simplicity, let $ $$B_{\pi}=\orp r{\pi}(i)$, $B_{\sigma}=\orp r{\sigma}(i)$,
and $B_{\omega}=\orp r{\omega}(i)$. We then have 
\begin{align*}
\left|B_{\sigma}\cap B_{\omega}\right| & =\left|B_{\sigma}\cap B_{\omega}\cap B_{\pi}^{c}\right|+\left|B_{\sigma}\cap B_{\omega}\cap B_{\pi}\right|\\
 & \le\left|B_{\omega}\cap B_{\pi}^{c}\right|+\left|B_{\sigma}\cap B_{\pi}\right|\\
 & \stackrel{\mathsf{(a)}}{<}\left(r-\left|B_{\omega}\cap B_{\pi}\right|\right)+\left(r-2t\right)\\
 & \stackrel{(\mathsf{b})}{\le}t+r-2t=r-t,
\end{align*}
where $B_{\pi}^{c}$ denotes the complement of $B_{\pi}$. Inequality
$\mathsf{(a)}$ follows from the fact that $\left|B_{\omega}\right|=r$
and~(\ref{eq:construct-1}); and inequality $\mathsf{(b)}$ follows
from~(\ref{eq:t-errors}). This completes the proof.\end{IEEEproof}
\begin{rem}
\label{rem:gen-errors}A code satisfying the condition of Lemma~\ref{lem:construct-1}
can in fact correct a class of errors that is more general than translocation
errors. More precisely, the code can correct errors that lead to the
displacement of at most $t$ elements of each rank. In particular,
the code can correct $t$ transposition errors, $t$ Hamming errors,
or any $t$ errors where each error displaces at most one element
from each rank. As an example of the latter type of error, consider
\begin{multline*}
\left(\{3,\mathbf{4}\},\{2,6\},\{\mathbf{7},8\},\{\mathbf{1},5\}\right)\xrightarrow{\mbox{ error }}\\
\left(\{3,\mathbf{1}\},\{2,6\},\{\mathbf{4},8\},\{\mathbf{7},5\}\right),
\end{multline*}
where each rank corresponds to one set in the set partition. We note
that each part except for the one listed second has one displaced
(moved) element.
\end{rem}
A code that satisfies the conditions of Lemma~\ref{lem:construct-1}
can be decoded in time $O(Mnr)$, where $M$ is the size of the code.
As before, suppose that $\pi\in C$ is the unknown stored codeword
and $\omega$ is the retrieved permutation. For each $i\in[n/r]$,
we must identify a unique set $A(i)\in\{\orp r{\sigma}(i):\sigma\in C\}$
such that 
\begin{equation}
\left|A(i)\cap\orp r{\omega}(i)\right|\ge r-t.\label{eq:decoding-intersection}
\end{equation}
The ordered partition representation of $\pi$ is then $\orp r{\pi}=\left(A(1),\dotsc,A(n/r)\right).$

The intersection of $\orp r{\omega}(i)$ and each of the sets in $\{\orp r{\sigma}(i):\sigma\in C\}$
can be trivially found with time complexity $O(r^{2})$. Since there
are $M$ sets in $\{\orp r{\sigma}(i):\sigma\in C\}$, finding $A(i)$
for each $i\in[n/r]$ takes $O(Mr^{2})$ steps. Thus $\orp r{\pi}$
can be identified with complexity $O(Mr^{2}n/r)=O(Mnr)$. 

Since for some code parameters $M$ can be exponential in $n$, the
time needed for exhaustive search decoding may be exponential as well.
However, if more information about the structure of the code is available,
decoding may be performed much faster, as in the cases of constructions
based on grouping elements and Steiner systems discussed in Subsections~\ref{sub:grouping}
and \ref{sub:designs}.

We pause to briefly comment on the relationship between almost disjoint
sets of set partitions and intersecting families, in the context of
the celebrated Erd\H{o}s-Ko-Rado (EKR) theorem (see~\cite{Deza1983EKR22years,Ellis2012Setwise}
and references therein). A family of subsets of a set is said to be
intersecting if each pair of subsets have a non-empty intersection.
The EKR theorem establishes upper bounds on the size of the largest
intersecting family. This theorem is also extended to the space of
permutations where a set of permutations is said to be intersecting
if each pair of permutations agree in some coordinate~\cite{frankl-deza,cameron2003intersecting,Godsil2009EKR}.
In our formulation, we require the intersections to be small, unlike
for intersecting families where the intersection size may be arbitrary
large as long as it is non-zero. Furthermore, we require our subsets
to be organized into ordered partitions, with the intersection property
holding only for parts at the same location. Although the code-anticode
theorem by Delsarte~\cite{delsarte1973algebraic,Schwartz2011Optimalpermutationanticodes}
may help in establishing bounds on families of subsets intersecting
in a few elements only, it cannot be used for the specialized ordered
set partition setting in a simple manner. To the best of our knowledge,
the almost disjoint set partition family problem has not been previously
studied in the extremal combinatorics literature.

Next, we describe two methods for constructing codes that satisfy
the conditions of Lemma~\ref{lem:construct-1}.

\subsubsection{A Construction based on grouping elements\label{sub:grouping}}

If $r$ is a multiple of $2t+1$, the following simple construction
satisfies the conditions of Lemma~\ref{lem:construct-1}. Partition
the set $\left[n\right]$ in an arbitrary fashion into $n/(2t+1)$
parts $E_{1},\cdots,E_{n/(2t+1)}$, each of size $2t+1$. Consider
all ordered partitions $\op$ of $[n]$ into $n/r$ parts of size
$r$ that place all elements of each $E_{j},j\in[n/(2t+1)],$ in the
same part. Let $C$ be a code such that its corresponding set of ordered
set partitions $\Op_{r}(C)$ consists of the set of aforementioned
partitions $\op$.

As an illustration, suppose $t=1$, $r=6$, and $n=12$, and let $\{1,\dotsc,12\}$
be partitioned as $\{E_{1},E_{2},E_{3},E_{4}\}$, with 
\begin{align*}
E_{1} & =\{1,2,3\}, & E_{2} & =\{4,5,6\},\\
E_{3} & =\{7,8,9\}, & E_{4} & =\{10,11,12\}.
\end{align*}
Next, consider ordered partitions of $\{1,\dotsc,12\}$ that place
all elements of each $E_{i}$ in the same part, namely,
\begin{align*}
\op_{1} & =\left(\{\underline{1,2,3},\underline{4,5,6}\},\{\underline{7,8,9},\underline{10,11,12}\}\right),\\
\op_{2} & =\left(\{\underline{1,2,3},\underline{7,8,9}\},\{\underline{4,5,6},\underline{10,11,12}\}\right),\\
\op_{3} & =\left(\{\underline{1,2,3},\underline{10,11,12}\},\{\underline{4,5,6},\underline{7,8,9}\}\right),\\
\op_{4} & =\left(\{\underline{4,5,6},\underline{7,8,9}\},\{\underline{1,2,3},\underline{10,11,12}\}\right),\\
\op_{5} & =\left(\{\underline{4,5,6},\underline{10,11,12}\},\{\underline{1,2,3},\underline{7,8,9}\}\right),\\
\op_{6} & =\left(\{\underline{7,8,9},\underline{10,11,12}\},\{\underline{1,2,3},\underline{4,5,6}\}\right).
\end{align*}
Then, the code corresponding to the set of ordered partitions $\Op=\{\op_{1},\dotsc,\op_{6}\}$
can correct one translocation error.

To see that $C$ satisfies the conditions of Lemma~\ref{lem:construct-1},
consider $\op,\op'\in\Orp rC$ and $i\in[n/r]$. Suppose that $\op(i)\neq\op'(i)$.
There exists $E_{j}$ such that $E_{j}\subseteq\op(i)$ but $E_{j}\cap\op'(i)=\emptyset$.
Since $|E_{j}|=2t+1$, we have $\left|\op\cap\op'(i)\right|<r-2t$.

The simplicity of this construction allows for fast decoding. Without
loss of generality, assume that 
\[
E_{j}=\{(j-1)(2t+1)+1,\dots,j(2t+1)\}.
\]
Suppose that $\pi$ is the stored codeword and $\omega$ is the retrieved
permutation. For each $i\in[n/r]$, we have $E_{j}\subseteq\orp r{\pi}(i)$
if $|E_{j}\cap\orp r{\omega}(i)|\ge t+1$. To compute $|E_{j}\cap\orp r{\omega}(i)|$,
$j\in\left[n/\left(2t+1\right)\right]$, we compare each element of
$\orp r{\omega}(i)$ with $j\left(2t+1\right)$, $j\in\left[n/\left(2t+1\right)\right]$.
This can be performed in $O\left(\frac{rn}{2t+1}\right)$ steps. Hence,
decoding can be performed in time $O(\frac{n}{r}\frac{rn}{2t+1})=O(n^{2})$.

Let $d=2t+1.$ The cardinality of the code $C$ equals 
\[
\frac{(n/d)!}{\left((r/d)!\right)^{n/r}},
\]
and thus the asymptotic rate is 
\begin{align*}
\lim\frac{\frac{n}{d}\ln\frac{n}{d}-\frac{n}{d}\ln\frac{r}{d}+O(n)}{n\ln n+O(n)} & =\lim\frac{1}{d}\frac{\ln n-\ln r+O(1)}{\ln n+O(1)}\\
 & =(1-\rho)\lim\frac{1}{d}.
\end{align*}
Hence, the asymptotic rate is nonzero iff $d$ is bounded (constant).
While the rate of the code does not approach capacity, it should be
noted that, per Remark~\ref{rem:gen-errors}, the code can correct
more general errors than translocation errors.

\subsubsection{Constructions based on combinatorial designs\label{sub:designs}}

Several well-known -- and a number of significantly lesser known --
families of combinatorial objects are closely related to the notion
of almost disjoint ordered set partition families. These include block
designs and Latin squares. From the first category, we use \emph{Steiner
systems }and \emph{resolvable balanced incomplete block designs} and,
from the latter category, we mention \emph{semi-Latin squares}, representing
a generalization of the well-known family of Latin squares \cite{stinson2004combinatorial}.
The constructions are straightforward consequences of the definition
of almost disjoint sets, but they provide for a rather limited set
of code parameters. A more general method, based on interleaving arguments,
will be presented in the next subsection.

A Latin square of order $n$ is an $n\times n$ array such that each
element of $[n]$ appears exactly once in each row and exactly once
in each column. A semi-Latin square with parameters $n$ and $r$
is an $\frac{n}{r}\times\frac{n}{r}$, array where each cell is an
$r$-subset of $[n]$ such that each element in $[n]$ appears exactly
once in each column and exactly once in each row~\cite{Preece1983SemiLatin}.
An example of a semi-Latin square is shown below, with $n=6,\, r=2$:

\begin{table}[H]
\noindent \centering{}%
\begin{tabular}{|c|c|c|}
\hline 
\{1,4\}  & \{2,5\}  & \{3,6\} \tabularnewline
\hline 
\{3,5\}  & \{1,6\}  & \{2,4\} \tabularnewline
\hline 
\{2,6\}  & \{3,4\}  & \{1,5\} \tabularnewline
\hline 
\end{tabular}
\end{table}

Note that the definition of a semi-Latin square implies that each
row and each column of the square represent a partition of $[n]$.
Hence, we arrive at the following result.
\begin{lem}
\label{lem:semi-latin-code}The rows of a semi-Latin square with parameters
$n$ and $r$, viewed as ordered set partitions of $[n]$, form the
ordered set partitions of an $\mpcode{\circ}(n,r,r)$ code of cardinality
$\frac{n}{r}$.
\end{lem}
The result is a direct consequence of Lemma~\ref{lem:construct-1}
and the fact that no element is repeated in a column of a semi-Latin
square. Unfortunately, the size of a code based on semi-Latin squares
is small, since the row-column restrictions are too strong for the
purpose of designing almost disjoint ordered set partition families. 

As stated before, a code that satisfies the conditions of Lemma~\ref{lem:construct-1}
can be decoded in time $O(Mnr)$. This implies that the code of Lemma~\ref{lem:semi-latin-code}
is decodable in time $O(n^{2})$.

Another family of combinatorial objects that allow for constructing
almost disjoint ordered set partitions are special types of designs,
namely \emph{resolvable balanced incomplete block designs} and \emph{resolvable
Steiner systems}.

A $k$-$(n,r,\lambda)$\emph{-design} is a family of $r$-subsets
of a set $X$ of size $n$, each called a \emph{block}, such that
every $k$-subset of $X$ appears in exactly $\lambda$ blocks. Such
a design is \emph{resolvable }if its blocks can be grouped into $m$
classes, such that each class forms a partition of $X$. It is known
that~\cite[p.~202]{stinson2004combinatorial}
\[
m=\lambda\frac{\binom{n-1}{k-1}}{\binom{r-1}{k-1}}\cdot
\]

A \emph{Steiner system} $S(k,r,n)$ is a $k$-$(n,r,1)$-design and
a \emph{balanced incomplete block design} (BIBD) with parameters $(n,r,\lambda)$
is a $2$-$(n,r,\lambda)$-design. For the purpose of code construction,
resolvable Steiner systems and Resolvable BIBDs (RBIBDs) are of special
interest.

The following lemma shows that resolvable Steiner systems can be used
to construct multipermutation codes in the Ulam metric. Resolvable
designs may also be used to construct multipermutation codes in the
Hamming metric, as described by Chu et al.~\cite{Chu2006OnCCCs}.
The aforementioned construction nevertheless does not cater to the
specialized requirements posed by the Ulam metric.
\begin{lem}
\label{lem:k-designs}If a resolvable Steiner system $S(k,r,n)$ exists,
then there exists an $\mpcode{\circ}(n,r,d)$, where $d$ is an odd
number satisfying $d\le r-k+1$, of size 
\[
\frac{\binom{n-1}{k-1}}{\binom{r-1}{k-1}}\left(\frac{n}{r}\right)!.
\]
\end{lem}
\begin{IEEEproof}
We use a Steiner system $S(k,r,n)$ to construct a family of ordered
set partitions satisfying the conditions of Lemma~\ref{lem:construct-1}.

Let $m$ denote the number of classes of the Steiner system. The blocks
of each of the $m$ classes of the Steiner system form an unordered
set partition. Each unordered set partition gives rise to $\left(\frac{n}{r}\right)!$
ordered set partitions. Hence, in total, we have $m(\frac{n}{r})!$
ordered set partitions. Let $C$ be a code such that its corresponding
set of ordered set partitions $\Orp rC$ is the aforementioned set
of $m(\frac{n}{r})!$ partitions. 

Let $t=(d-1)/2$. We have $k\le r-2t$. Each two blocks in the Steiner
system have less than $k$ elements in common, and consequently, have
less than $r-2t$ elements in common. It follows that the conditions
of Lemma~\ref{lem:construct-1} are satisfied. Hence, $C$ is an
$\mpcode{\circ}(n,r,d)$ of the stated size.
\end{IEEEproof}
The code described in the preceding lemma can be decoded in time $O(n^{k}r)$
as follows. Suppose that $\pi$ is the stored codeword and $\omega$
is the retrieved permutation. For each $i\in[n/r]$, to find $\orp r{\pi}(i)$,
one needs to compute the size of the intersection of $\orp r{\omega}(i)$
with the blocks of the Steiner system. Computing each intersection
takes $O(r^{2})$. Hence, decoding can be performed in time 
\[
O\left(\frac{n}{r}\frac{\binom{n-1}{k-1}}{\binom{r-1}{k-1}}r^{2}\right)=O\left(nr\binom{n-1}{k-1}\right)=O\left(n^{k}r\right).
\]

An RBIBD with parameters $(n,r,\lambda=1)$ is a resolvable Steiner
system $S(2,r,n)$, and thus can be used for code construction. For
$\lambda=1$, the case of interest in all our subsequent derivations,
the condition 
\[
n=r\;\mod r(r-1)
\]
is necessary for the existence of an RBIBD, and it is also known to
be asymptotically sufficient for $r\geq5$~\cite{rumov1976existence}.

Two of the most commonly used approaches to constructing RBIBDs are
based on finite fields~\cite{rumov1976existence} and on a simple
combinatorial construction~\cite{khare1981simple}. Using the former
construction, one can derive RBIBDs with parameters $\lambda=1,$
$n=p^{\alpha v}$, and $r=p^{\alpha}$, with $p$ a prime and $\alpha$
and $v$ positive integers. 

The combinatorial construction of~\cite{khare1981simple} is based
on the following straightforward procedure. Assume that $r$ is prime
and arrange the $n=r^{2}$ elements of the $n$-set into an $r\times r$
array in order. Each row corresponds to one block of size $r$, and
each array represents a class that partitions the $n$-set. The first
class, denoted by $C_{1}$, is shown below for $r=3$:
\begin{table}[H]
\centering{}%
\begin{tabular}{|c|c|c|}
\hline 
1  & 2  & 3 \tabularnewline
\hline 
4  & 5  & 6 \tabularnewline
\hline 
7  & 8  & 9 \tabularnewline
\hline 
\end{tabular}
\end{table}

Class $C_{2}$ is constructed from class $C_{1}$ by taking the transpose.
Each subsequent class $C_{i}$, for $i\geq3$, is constructed from
the previous class $C_{i-1}$ in the following manner: the cyclically
continued diagonals of $C_{i-1}$ are arranged row-wise, starting
from the main diagonal, and then moving to the left sub-diagonals.
For the example with $r=3$, the additional three classes constructed
according to the above procedure take the form:
\begin{table}[H]
\centering{}%
\begin{tabular}{|c|c|c|}
\hline 
1  & 4  & 7 \tabularnewline
\hline 
2  & 5  & 8 \tabularnewline
\hline 
3  & 6  & 9 \tabularnewline
\hline 
\end{tabular}~~~~%
\begin{tabular}{|c|c|c|}
\hline 
1  & 5  & 9 \tabularnewline
\hline 
2  & 6  & 7 \tabularnewline
\hline 
3  & 4  & 8 \tabularnewline
\hline 
\end{tabular}~~~~%
\begin{tabular}{|c|c|c|}
\hline 
1  & 6  & 8 \tabularnewline
\hline 
2  & 4  & 9 \tabularnewline
\hline 
3  & 5  & 7 \tabularnewline
\hline 
\end{tabular} 
\end{table}

Note that the procedure terminates after $r+1$ steps, resulting in
a repetition of class $C_{2}$. The total number of blocks in the
RBIBD equals $r(r+1)=r^{2}+r$. 

The construction involving cyclic diagonal shifts can be extended
for resolvable, \emph{unbalanced} IBDs with parameters $n=p^{\alpha}r$,
$\alpha\geq1$, $p$ prime, and block size $r$ which may be an arbitrary
integer $\geq2$. The only difference between a balanced and unbalanced
design is the requirement that any pair of elements appear in \emph{at
most} $\lambda$ blocks\ \cite{khare1981simple}. For the case $\lambda=1,$
i.e., any pair of elements appearing zero or one time, the designs
are known as zero-one concurrence designs; they may be constructed
by a combination of variety cutting and the diagonalization procedure
described above. The interested reader is referred to \cite{khare1981simple}
for an in-depth treatment of this construction. 

The aforementioned procedures show that an RBIBD with parameters $(r^{2},r,1)$
exists, provided that $r$ is am odd prime. Hence, using Lemma \ref{lem:k-designs},
we can obtain the following lemma, which concludes this subsection.
\begin{lem}
Suppose that $r$ is an odd prime. Then, $A_{\circ}(r^{2},r,r-2)\ge(r+1)r!$.
\end{lem}

\subsection{Construction based on  codes with $r$ components \label{sub:Ulam-rcomps}}

In this subsection, we present a construction for multipermutation
codes in the Ulam metric based on $r$ permutation codes of length
$n/r$, interleaved to ensure translocation error protection.

Assume first that $d\le n/r$. Consider a partition $\{P_{1},\dotsc,P_{r}\}$
of $[n]$ into sets of equal size, and the set of codes $\{C_{1},\dotsc,C_{r}\}$,
with each $C_{i},i\in[r],$ being a permutation code of minimum Ulam
distance $d$ over $P_{i}$. We form a new code $C$ as follows:
\begin{equation}
C=\bigcup_{c_{i}\in C_{i},\,\forall\, i}\eqclass r(c_{1}\circ\dotsm\circ c_{r}).\label{eq:interleaved-code}
\end{equation}

\begin{prop}
\label{prop:rcomps}The code $C$ given in~(\ref{eq:interleaved-code})
is an $\mpcode{\circ}(n,r,d)$ code.\end{prop}
\begin{IEEEproof}
Consider $\pi',\sigma'\in C$ such that $\pi'\noteqv r\sigma'$. By
construction, there exists $\pi\in\eqclass r(\pi')$ and $\sigma\in\eqclass r(\sigma')$
such that
\begin{align}
\sigma & =\sigma_{1}\circ\dotsm\circ\sigma_{r},\quad\sigma_{i}\in C_{i},\nonumber \\
\pi & =\pi_{1}\circ\dotsm\circ\pi_{r},\quad\pi_{i}\in C_{i}.\label{eq:rcomps1}
\end{align}
Since $\pi\noteqv r\sigma$, there exists an element $j\in[r]$ such
that $\pi_{j}\neq\sigma_{j}$. 

We show that for an arbitrary choice of $\alpha\in\eqclass r(\pi)$
and $\beta\in\eqclass r(\sigma)$, we have $\tdist{}(\alpha,\beta)\ge d$.
Since $\alpha$ and $\beta$ are chosen arbitrarily, we find that
\[
\tdist r(\pi',\sigma')=\tdist r(\pi,\sigma)=\min_{\alpha\in\eqclass r(\pi)}\min_{\beta\in\eqclass r(\sigma)}\tdist{}(\alpha,\beta)\ge d,
\]
which completes the proof. 

For $\alpha\in\eqclass r(\pi)$ and $\beta\in\eqclass r(\sigma)$
and each $i\in[r]$, the order of the elements of $P_{i}$ is the
same in $\pi$ and in $\alpha$, i.e., $\proj{\pi}{P_{i}}=\proj{\alpha}{P_{i}}.$
Furthermore, since $\pi_{i}\in C_{i}$, and $C_{i}$ is a code over
$P_{i}$, we have $\pi_{i}=\proj{\pi}{P_{i}}$. Hence, $\proj{\alpha}{P_{i}}=\pi_{i}$.
A similar argument holds for $\sigma$ and $\beta$, implying that
$\proj{\beta}{P_{i}}=\sigma_{i}$ for each $i\in[r]$. So, by Lemma~\ref{lem:projection},
one can show that
\begin{align*}
\tdist{}(\alpha,\beta) & \ge\sum_{i=1}^{r}\tdist{}\left(\proj{\alpha}{P_{i}},\proj{\beta}{P_{i}}\right)=\sum_{i=1}^{r}\tdist{}\left(\pi_{i},\sigma_{i}\right)\\
 & \ge\tdist{}\left(\pi_{j},\sigma_{j}\right)\ge d,
\end{align*}
where the last inequality follows from $\pi_{j}\neq\sigma_{j}$.
\end{IEEEproof}
As an example, for $n=6$, $r=2$, and $d=2$, consider
\begin{align*}
P_{1} & =\left\{ 1,2,3\right\} ,\\
P_{2} & =\left\{ 4,5,6\right\} ,\\
C_{1} & =\left\{ \left(1,2,3\right),\left(3,2,1\right))\right\} ,\\
C_{2} & =\left\{ \left(4,5,6\right),\left(6,5,4\right)\right\} .
\end{align*}
Note that $C_{1}$ and $C_{2}$ both have Ulam distance equal to 2.
The code $C$, constructed according to~(\ref{eq:interleaved-code})
contains
\[
\begin{array}{cc}
\left(1,4,2,5,3,6\right), & \left(1,6,2,5,3,4\right),\\
\left(3,4,2,5,1,6\right), & \left(3,6,2,5,1,4\right),
\end{array}
\]
and their equivalency classes under $\eqv 2$. For instance, let $\pi=\left(1,4,2,5,3,6\right)$
and $\sigma=\left(3,4,2,5,1,6\right)$ and consider $\alpha=\left(4,1,5,2,3,6\right)\in\eqclass 2\left(\pi\right)$
and $\beta=\left(4,3,5,2,6,1\right)\in\eqclass 2\left(\sigma\right)$.
It can be observed that $\alpha_{P_{1}}=\pi_{P_{1}}=\left(1,2,3\right)$
and $\alpha_{P_{2}}=\pi_{P_{2}}=\left(4,5,6\right)$. Similar statements
hold for $\beta$ and $\sigma$. It can also be verified that 
\[
\tdist{}\left(\alpha,\beta\right)=2.
\]

For several constructions of permutation codes in the Ulam metric,
we refer the reader to~\cite{farnoud2013error-correction}. 

The components of the constructed code can be decoded independently.
As before, suppose that $\pi$ is the stored codeword and $\omega$
is the retrieved permutation. Since there are at most $t=\left\lfloor \frac{d-1}{2}\right\rfloor $
errors, we have $\tdist{}(\pi,\omega)\le t$. By Lemma~\ref{lem:projection},
this implies that $\tdist{}(\pi_{P},\omega_{P})\le t$ for all $P\in\{P_{1},\dotsc,P_{r}\}$.
Hence, one can use a decoder for permutation codes in the Ulam metric
that can correct $t$ errors. Consequently, $\orp r{\pi}$ can be
identified from $\omega_{P},P\in\{P_{1},\dotsc,P_{r}\},$ through
a parallel decoding process. Note that a simple decoding architecture
for a class of codes in the Ulam metric was proposed in our companion
paper~\cite{farnoud2013error-correction}, based on Hamming distance
decoding of de-interleaved component codes.

Assuming that the cardinality of the codes $C_{i}$ equals $A_{\circ}(n/r,1,d)$,
the cardinality of $C$ equals $A_{\circ}(n/r,1,d)^{r}.$ Recall that
we define the cardinality of a multipermutation code as the number
of its equivalency classes and not the number of its elements. It
was proved in~\cite{farnoud2013error-correction} that 
\[
A_{\circ}(m,1,d)\ge\frac{(m-d+1)!}{\binom{m}{d-1}}\cdot
\]
Hence,
\[
A_{\circ}(n,r,d)\ge\left(\frac{(n/r-d+1)!}{\binom{n/r}{d-1}}\right)^{r}.
\]
Furthermore, from the fact that $\mathcal{C}_{\circ}(1,d)=1-\delta$~\cite{farnoud2013error-correction},
we find that 
\begin{align*}
\mathcal{C}_{\circ}(r,d) & =\lim\frac{\ln A_{\circ}(n,r,d)}{\ln n!}\\
 & \ge\lim\frac{r\ln A_{\circ}(n/r,1,d)}{\ln n!}\\
 & =\lim\frac{\ln A_{\circ}(n/r,1,d)}{\ln(n/r)!}\lim\frac{r\ln(n/r)!}{\ln n!}\\
 & =(1-\lim\frac{rd}{n})(1-\rho).
\end{align*}
In particular, if $\lim\frac{rd}{n}=0$, then $\mathcal{C}_{\circ}(r,d)=(1-\rho).$

\subsection{Construction based on codes in the Hamming metric}

Recall that $\hdist r(\pi,\sigma)\ge\tdist r(\pi,\sigma).$ Thus,
if $C$ is an $\mpcode{\circ}(n,r,d)$ code, then it is also an $\mpcode H(n,r,d)$
code. We now show that an $\mpcode{\circ}(n,r,d)$ code can be obtained
using multipermutation Hamming codes of shorter lengths. We refer
the reader to~\cite{Luo2003CCCZq,Ding2005CombinatorialCCC,Huczynska2006FrequencyPermutationArrays,Chu2006OnCCCs}
for constructions of multipermutation codes in the Hamming metric.
\begin{prop}
\label{prop:one-stage}Suppose that $n/r$ is even and that $d\le r$.
Let $P=\left[\frac{n}{2}\right]$, and $Q=[n]\backslash P$. Additionally,
let $C'_{1}$ be an $\mpcode{\circ}(\frac{n}{2},r,d)$ code over $P$
and $C_{1}$ be an $\mpcode H(\frac{n}{2},r,d)$ code over $Q$. The
code $C=C'_{1}\circ_{r}C_{1}$ is an $\mpcode{\circ}(n,r,d)$ code.\end{prop}
\begin{IEEEproof}
Let $\pi,\sigma\in C$ with $\pi\noteqv r\sigma$. Assume that 
\begin{align*}
\pi & =\pi'_{1}\circ_{r}\pi_{1}, & \sigma & =\sigma'_{1}\circ_{r}\sigma_{1},
\end{align*}
where $\pi_{1}',\sigma_{1}'\in C_{1}'$ and $\pi{}_{1},\sigma_{1}\in C_{1}$. 

First, suppose that $\pi_{1}'\noteqv r\sigma_{1}'$. Then,
\begin{align*}
\tdist r(\pi,\sigma) & =\min_{\alpha\in\eqclass r(\pi)}\min_{\beta\in\eqclass r(\sigma)}\tdist{}(\alpha,\beta)\\
 & \ge\min_{\alpha\in\eqclass r(\pi)}\min_{\beta\in\eqclass r(\sigma)}\tdist{}(\alpha_{P},\beta_{P})\\
 & \ge d,
\end{align*}
where the first inequality follows from Lemma~\ref{lem:projection},
and the second inequality follows from the facts that $\alpha_{P}\in\eqclass r(\pi_{1}')\subseteq C_{1}'$,
$\beta_{P}\in\eqclass r(\sigma_{1}')\subseteq C_{1}'$, and that $C_{1}'$
is an $\mpcode{\circ}(n/2,r,d)$ code.

Next, suppose that $\pi_{1}'\eqv r\sigma_{1}'$. Since $\pi\noteqv r\sigma$,
we have $\pi_{1}\noteqv r\sigma_{1}$. Let 
\[
D=\left\{ x\in Q:x\in\orp r{\pi_{1}}(i),x\in\orp r{\sigma_{1}}(j),i\neq j\right\} 
\]
be the set of elements of $Q$ that are of different ranks in $\pi_{1}$
and $\sigma_{1}$. Note that $|D|=\hdist r(\pi_{1},\sigma_{1})$.

Consider $\alpha\in\eqclass r(\pi)$ and $\beta\in\eqclass r(\sigma)$.
For odd values of $i$, we have $\orp r{\alpha}(i)=\orp r{\beta}(i)$,
as $\pi_{1}'\eqv r\sigma_{1}'$. 

On the one hand, for any common subsequence of $\alpha$ and $\beta$
that contains an element of $D$, there exists some odd $i$ such
that $\orp r{\alpha}(i)=\orp r{\beta}(i)$ is not in that subsequence.
This implies that the length of the given common subsequence is at
most $n-r$. On the other hand, for any common subsequence of $\alpha$
and $\beta$ that does not contain any element of $D$, the length
of that subsequence is at most 
\begin{align*}
n-|D| & =n-\hdist r(\pi_{1},\sigma_{1})\le n-d.
\end{align*}
Hence, the length of any common subsequence of $\alpha$ and $\beta$
is at most 
\[
\max\{n-d,n-r\}=n-d
\]
and thus $\tdist{}(\alpha,\beta)\ge d.$ Since $\alpha$ and $\beta$
are arbitrary elements of $\eqclass r(\pi)$ and $\eqclass r(\sigma)$,
respectively, we find that $\tdist r(\pi,\sigma)\ge d,$ which completes
the proof.
\end{IEEEproof}
One particularly simple choice for $C_{1}'$ is 
\begin{equation}
C_{1}'=\eqclass r\left((1,\dotsc,n/2)\right),\label{eq:one-codeword}
\end{equation}
which is a code with cardinality 1. 

As an example, let $n=8$, $r=2$, $d=4$, and 
\begin{align*}
C_{1}' & =\eqclass 2\left(\left(1,2,3,4\right)\right)\\
 & =\left\{ \left(1,2,3,4\right),\left(2,1,3,4\right),\left(1,2,4,3\right),\left(2,1,4,3\right)\right\} ,\\
C_{1} & =\eqclass 2\left(\left(5,6,7,8\right)\right)\cup\eqclass 2\left(\left(7,8,5,6\right)\right),
\end{align*}
which leads to
\[
C=\eqclass 2\left(\left(1,2,5,6,3,4,7,8\right)\right)\cup\eqclass 2\left(\left(1,2,7,8,3,4,5,6\right)\right),
\]
an $\mpcode{\circ}\left(8,2,4\right)$ code. 

For the case of~(\ref{eq:one-codeword}), the cardinality of $C$
equals the cardinality of $C_{1}$, which may be as large as $A_{H}(n/2,r,d).$
Hence,
\[
A_{\circ}(n,r,d)\ge A_{H}(n/2,r,d)
\]
 if $n/r$ is even and $d\le r$. With similar arguments, one can
show that, if $n/r$ is odd and $d\le r$, then 
\[
A_{\circ}(n,r,d)\ge A_{H}\left((n+r)/2,r,d\right).
\]
For $d\le r$ and $\rho<1$, we have $\delta=0$. Hence, for $d\le r$
and $\rho<1$, 
\[
\mathcal{C}_{\circ}(r,d)\ge\frac{1}{2}(1-\rho)(1-2\delta)=\frac{1}{2}(1-\rho).
\]

To construct larger codebooks, one may recursively use the construction
of Prop.~\ref{prop:one-stage} to design $C_{1}'$. For simplicity,
suppose that $n$ and $r$ are both powers of 2. Let 
\[
C=\left(\left(C'_{k}\circ_{r}C_{k}\right)\circ_{r}C_{k-1}\circ_{r}\dotsm\right)\circ_{r}C_{1},
\]
where each $C_{i},i\in[k],$ is an $\mpcode H\left(n/2^{i},r,d\right)$
code, $C_{k}'=\eqclass r\left((1,\dotsc,n/2^{k})\right)$, and $k$
is a positive integer satisfying $k\le\lg(n/r)$. The condition $k\le\lg(n/r)$
is required since we need $n/2^{k}\ge r$. Note that this condition
also implies that $n/2^{k}\ge d$, since $d\le r$. The code $C$
is an $\mpcode{\circ}(n,r,d)$ code. The cardinality of $C_{i}$ can
be as large as $A_{H}\left(n/2^{i},r,d\right)$. Hence, if $n$ and
$r$ are powers of 2 and $d\le r$, it holds that
\[
A_{\circ}(n,r,d)\ge\prod_{i=1}^{k}A_{H}\left(n/2^{i},r,d\right).
\]
Let $n=2^{j}$, $r=2^{\rho j}$, $d\le r$, where $\rho$ is a constant
less than 1, and suppose that $k$ is a constant such that $k\le\lg(n/r)=j(1-\rho)$.
For this regime, we have
\begin{align*}
\lim_{j\to\infty}\frac{\ln A_{\circ}(2^{j},2^{\rho j},d)}{\ln n!} & \ge\lim_{j\to\infty}\frac{\lg A_{\circ}(2^{j},2^{\rho j},2^{\rho j})}{\lg2^{j}!}\\
\ge & \sum_{i=1}^{k}\lim_{j\to\infty}\frac{\lg A_{H}(2^{j-i},2^{\rho j},2^{\rho j})}{\lg2^{j-i}!}\frac{\lg2^{j-i}!}{\lg2^{j}!}\\
= & \sum_{i=1}^{k}\left(1-\lim_{j\to\infty}\frac{\rho j}{j-i}\right)2^{-i}\\
= & (1-\rho)(1-2^{-k}).
\end{align*}
Since $\rho<1$, $k$ can be chosen arbitrarily large. Hence, the
asymptotic rate can be made arbitrary close to $(1-\rho)$.

\section{Conclusion\label{sec:Conclusion}}

We studied a novel rank modulation scheme based on multipermutation
codes in the Ulam metric. We also highlighted the close connection
between multipermutation codes in the Hamming metric, also known as
constant composition codes and frequency permutation arrays, and codes
in the Ulam metric.

The presented results included bounds on the size of multipermutation
codes in both the Ulam metric and the Hamming metric; for the case
of the Hamming metric, these bounds led to the capacity of the codes,
while for the Ulam metric, the bounds led to upper bounds and lower
bounds for the capacity, with a gap equal to $\rho(1-\delta)$. We
also presented several construction methods for codes in the Ulam
metric using permutation interleaving, semi-Latin squares, resolvable
Steiner systems, and resolvable balanced incomplete block designs,
among other techniques.

\paragraph*{Acknowledgment}

The work was supported by NSF grants CCF 0809895, CCF 1218764, and
the Emerging Frontiers for Science of Information, CCF 0939370. The
authors would like to thank anonymous reviewers for their insightful
comments and Eyal En Gad for several useful discussions.

\bibliographystyle{ieeetr}
\addcontentsline{toc}{section}{\refname}\bibliography{IEEEfull,bib}

\end{document}